\documentclass[english,aps,preprintnumbers,amsmath,amssymb,nofootinbib,twocolumn]{revtex4}

\usepackage[latin1]{inputenc}
\usepackage{graphicx}
\usepackage{color}

\usepackage{color}

\usepackage{bm}
\usepackage{amsmath}

\usepackage{dsfont}

\newcommand{\be}{\begin{eqnarray}}
\newcommand{\ee}{\end{eqnarray}}

\newcommand{\nn}{\nonumber }

\newcommand{\Nf}{N_{\text{f}}}
\newcommand{\Nc}{N_{\text{c}}}
\newcommand{\pat}{\partial_t}
\newcommand{\Eqref}[1]{Eq.~\eqref{#1}}

\newcommand{\fslash}{\hspace{-0.1ex} \slash }
\def\slash#1{\setbox0=\hbox{$#1$}               
   \dimen0=\wd0                                 
   \setbox1=\hbox{/} \dimen1=\wd1               
   \ifdim\dimen0>\dimen1                        
      \rlap{\hbox to \dimen0{\hfil/\hfil}}      
      #1                                        
   \else                            
      \rlap{\hbox to \dimen1{\hfil$#1$\hfil}}   
      /                                         
   \fi}                                         %

\newcommand{\psibar}{\bar{\psi}}
\newcommand{\spinor}{\psi_{j}}
\newcommand{\barspinor}{\bar{\psi}_{j}}
\newcommand{\altbarspinor}{\bar{\psi}_{i}}
\newcommand{\altspinor}{\psi_{i}}

\newcommand{\wavpsi}{Z_{\psi}}

\usepackage{babel}
\makeatother

\begin{document}

\title{Asymptotic safety: a simple example}
\author{Jens Braun, Holger Gies and Daniel D.~Scherer}
\affiliation{Theoretisch-Physikalisches Institut, Friedrich-Schiller-Universit\"at Jena, 
Max-Wien-Platz 1, D-07743 Jena, Germany}

\begin{abstract}
  We use the Gross-Neveu model in $2<d<4$ as a simple fermionic example for
  Weinberg's asymptotic safety scenario: despite being perturbatively
  nonrenormalizable, the model defines an interacting quantum field theory
  being valid to arbitrarily high momentum scales owing to the existence of a
  non-Gau\ss ian fixed point. Using the functional renormalization group, we
  study the UV behavior of the model in both the purely fermionic as well as a
  partially bosonized language. We show that asymptotic safety is realized at
  non-Gau\ss ian fixed points in both formulations, the universal critical
  exponents of which we determine quantitatively. The partially bosonized
  formulation allows to make contact to the large-$\Nf$ expansion where the
  model is known to be renormalizable to all-orders. In this limit, the
  fixed-point action as well as all universal critical exponents can be
  computed analytically. As asymptotic safety has become an important scenario
  for quantizing gravity, our description of a well-understood 
  model is meant to provide for an easily accessible
  and controllable example of modern nonperturbative quantum field theory.
\end{abstract}

\maketitle

\section{Introduction}

Renormalizability is often described as a seemingly technical cornerstone for
the construction of admissible models in particle physics. Renormalization
fixes physical parameters of a model to measured values of observable
quantities. A prime physical meaning of renormalizability is the capability of a
model to provide an accurate description of a physical system over a wide
range of scales at which measurements can be performed. The set of
physical parameters, say, couplings or mass parameters etc.,  measured at
different scales then define the renormalized trajectory in parameter space
(theory space). If we demand for a
specific model to provide a {\em fundamental} description of nature, the model
must be valid on all scales, in particular down to arbitrarily short-distance
scales. The renormalized trajectory then must exist on all scales without
developing singularities. 

The requirement of renormalizability can formally be verified and realized in
perturbatively renormalizable theories in a weak coupling expansion. Here, all
free parameters of a model can be fixed to physical values and the
renormalized trajectory can be constructed order-by-order in a perturbative
expansion. This strategy can successfully be applied to theories that exist at
least over a wide range of scales, but still suffer from a maximum scale of UV
extension, such as QED~\cite{Landau:1955} or the standard
model Higgs sector~\cite{Brezin:1976bp}. If a
theory is asymptotically free, i.e., if the point of vanishing coupling
(Gau\ss ian fixed point) is a UV attractive fixed point, the perturbative
construction can even be applied on all scales, as in QCD.

Renormalizability is by no means bound to a perturbative construction. Even
though reliable nonperturbative information might be difficult to obtain, the
concept of renormalizability and the existence of a renormalized trajectory on
all scales can be formulated rather generally within Weinberg's asymptotic
safety scenario \cite{Weinberg:1976xy}. Loosely speaking, asymptotic safety is
the generalization of asymptotic freedom at the Gau\ss ian fixed point to the
case of a non-Gau\ss ian fixed point. As a fixed point of the renormalization
group (RG) by construction defines a point in parameter space where the system
becomes scale invariant, RG trajectories that hit the fixed point towards the
ultraviolet (UV) can be extended to arbitrarily high energy/momentum scales, thereby
defining a fundamental theory, for reviews see~\cite{Weinberg:1996kw,Rosten:2010vm}.

The asymptotic safety scenario has recently become an important ansatz for
quantizing gravity. In contrast to other approaches, this scenario is based on
the standard gravitational degrees of freedom and also the quantization
procedure proceeds in a rather standard fashion. Here, significant progress
was made with the aid of the functional RG, formulated in terms of a flow
equation~\cite{Wetterich:1992yh} for the effective average action for the
metric field~\cite{Reuter:1996cp}. In simple truncations, the RG flow of
gravity indeed reveals a non-Gau\ss ian fixed point
\cite{Dou:1997fg} -- a necessary prerequisite for
asymptotic safety. Most importantly, the fixed point has remained stable under
extensions of the truncation, {and its} universal properties such as the
critical exponents, in fact, exhibit a quantitative convergence under
improvements of the approximations involved
\cite{Lauscher:2001ya,Litim:2003vp,Codello:2006in}. RG
relevant directions in theory space have been identified and can be associated
with a finite number of physical parameters to be fixed by experiment. Taken
together, this provides for a rather strong evidence that a quantized version
of Einstein gravity can consistently be formulated within the asymptotic
safety scenario.

Still many questions are difficult to answer in the context of quantum
gravity, mainly due to technical and computational limitations. For a
confirmation of the asymptotic safety scenario, contact to other
nonperturbative quantization schemes have to be made in a quantitative manner;
first indications for a possible agreement, e.g., with dynamical
triangulations have already been observed \cite{Ambjorn:2005qt}.

UV-complete scenarios for the matter sector of the standard model built on
asymptotic safety have also been developed, for instance, for toy models of
the Higgs sector \cite{Gies:2003dp,Gies:2009hq,Gies:2009sv} also involving
non-linear sigma models \cite{Percacci:2009fh}, or for
extra dimensional Yang-Mills theories \cite{Gies:2003ic} and gravity
\cite{Litim:2003vp,Fischer:2006at}.

As asymptotic safety is inherently linked with field theories in the
nonperturbative domain, it appears highly worthwhile to identify and
investigate other nontrivial examples. The present work is devoted to such a
detailed study. Our benchmark model is given by the standard Gross-Neveu model
\cite{Gross:1974jv} in $d=3$ (or more generally in $2<d<4$) spacetime
dimensions. This model is known to be perturbatively nonrenormalizable as the
coupling constant carries a negative mass dimension. A naive loop expansion
leads to a series in terms of diagrams with an increasing superficial degree
of divergence. Proceeding in the standard fashion of perturbative
renormalization would require infinitely many counterterms and thus infinitely
many physical parameters to be fixed by experiment, implying that the theory
has no predictive power at all. In fact, it has long been known that this
conclusion is only an artifact of perturbative quantization. By means of a
Hubbard-Stratonovich transformation, the fermionic theory can be partially
bosonized such that an alternative expansion in terms of the inverse fermion
flavor number $\Nf$ can conveniently be formulated. The large-$\Nf$ expansion
turns out to be renormalizable to all orders rather similar to a small
coupling expansion in a perturbatively renormalizable model
\cite{Rosenstein:1988pt}. 

Whereas this provides strong indications for the existence of an interacting
Gross-Neveu model in $d=3$, it remains an open question as to whether this
conclusion holds for finite $\Nf$. On the other hand, one may wonder whether
this conclusion about nonperturbative renormalizability is indeed profoundly
nontrivial: as the partially bosonized version of the Gross-Neveu model is
identical to a Yukawa model. This seems to suggest that the renormalizability
in the large-$\Nf$ expansion may simply reflect the super-renormalizability of
the $d=3$ Yukawa model. {In fact, four-fermi models in $d=4$ are known to
  be related to Yukawa models near the Gau\ss ian fixed point
  \cite{ZinnJustin:1991yn,Hasenfratz:1991it}.} In this work, we wish to
emphasize that this is, in fact, not the case in $2<d<4$. As we show below,
also the bosonized Yukawa formulation is renormalized at a non-Gau\ss ian
fixed point within the asymptotic safety scenario. {Similar observations
  have been made from a more pragmatic viewpoint by studying the scaling
  properties of corresponding lattice models towards the continuum limit
  \cite{Hands:1992be,Karkkainen:1993ef}.}

This work mainly has a pedagogical character. Our analysis is performed in a
self-contained fashion within the modern formulation of the functional RG also
to provide guidance to the recent literature on asymptotically safe quantum
gravity. The fixed point analysis and the computation of universal properties
is performed explicitly and contact is made to the large-$\Nf$ expansion where
{fully} analytic results {for the fixed-point potential and all
  universal critical exponents} can be obtained.

As fermionic models occur in many circumstances of particle physics (effective
models of QCD) and many-body physics (strongly correlated electron systems),
our analysis of the microscopic completeness of the Gross-Neveu model also
provides a lesson for the functional RG treatment of such systems. {Of
  course, model studies of fermionic systems conventionally aim at long-range
  phenomena instead of short-distance behavior. As will be detailed below, the
  non-Gau\ss ian fixed-point facilitating UV asymptotic safety in the
  Gross-Neveu model at the same time serves as a quantum critical point
  associated with a 2nd order quantum phase transition towards a phase with
  broken discrete chiral symmetry. This phase transition has been studied
  already with a variety of techniques
  \cite{Hands:1992be,Karkkainen:1993ef,Gat:1991bf,Gracey:1993kc,Focht:1995ie,Hofling:2002hj}
  corresponding critical exponents are, of course, equivalent to those which
  we interpret as properties of the UV limit of the asymptotically safe
  model. }

Our work is organized as follows: {we begin with summarizing the essential
details of the model in Sect.~\ref{sec:GNm}. The basics of the asymptotic
safety scenario are summarized in Sect.~\ref{sec:AS}. An RG analysis in the
fermionic language is performed in Sect.~\ref{sec:fermionic}. Section
\ref{sec:bos} contains the corresponding study in the partially bosonized
formulation including a mean-field and large-$\Nf$ analysis and a numerical
evaluation of the functional RG equations.}

\section{Gross-Neveu model}
\label{sec:GNm}

The Gross-Neveu model describes the quantum field theory of $\Nf$ flavors of
massless relativistic fermions in $d$ space-time dimensions interacting via a
four-fermion interaction term. It allows to study dynamical chiral symmetry
breaking. The Euclidean action in $d$ space-time dimensions reads
\be
\label{eq:fermionic_action}
S[\psibar,\psi]
&=&\int_{x}\left\{\sum_{j=1}^{\Nf}\barspinor\mathrm{i}\fslash{\partial}\,\spinor
  + \sum_{i,j=1}^{\Nf}\altbarspinor\altspinor\frac{\bar{g}}{2
    \Nf}\barspinor\spinor\right\}\nn\\ 
&\equiv& \int_{x}\left\{\psibar\mathrm{i}\fslash{\partial}\,\psi+ \frac{\bar{g}}{2 \Nf}(\psibar\psi)^2
\right\}\,,
\ee
where $\int_{x}=\int d^{d} x$ is a shorthand for the integral over the
$d$-dimensional Euclidean space-time. The model depends on a single parameter
which is the coupling constant $\bar{g}$ {with} mass dimension $2-d$.

In this work we restrict ourselves to $2<d<4$; in $d=2$, the model is
asymptotically free and perturbatively renormalizable, as the Gau\ss ian fixed
point is UV attractive. $d=4$ will turn out to be a marginal case, where the
asymptotic safety scenario no longer applies {for integer $\Nf$} and the
theory becomes ``trivial'', i.e., non-interacting in the continuum limit, see
below and \cite{Gies:2009hq}.  To be specific, we employ a four-component
representation for the gamma matrices in the $d=3$ case in this work, i.e.,
$d_{\gamma}=4$ where $d_{\gamma}$ denotes the dimension of the Dirac algebra. In $d=3$, the explicit representation of our choice for the $4\times
4$ representation of the Dirac algebra can be written as
\be
\gamma_0=\tau_3\otimes\tau_3\,,\quad
\gamma_1=\tau_3\otimes\tau_1\,,\quad\gamma_2=\tau_3\otimes\tau_2\,.  
\ee
Here, the $\{\tau_i\}$'s denote the Pauli matrices which satisfy $\tau_i
\tau_j = \delta_{ij}\tau_0 + \mathrm{i}\epsilon_{ijk}\tau_k$, with
$i,j,k=1,2,3$ and $\tau_0=\mathds{1}_2$ is a $2\times2$ unit matrix. The
gamma matrices satisfy the anticommutation relation, i.e., the Dirac algebra
\be
\{\gamma_\mu,\gamma_\nu\}=2\delta_{\mu\nu}\mathds{1}_{4}\,, 
\ee where
$\mu,\nu=0,1,2$ and $\mathds{1}_{4}$ denotes the $4\times 4$ unit matrix.
Moreover we have two additional $4\times4$ matrices which anticommute with all
$\gamma_{\mu}$ and with each other: 
\be
\gamma_3=-\tau_{1}\otimes\tau_{0}\,,\quad\gamma_5=\tau_2\otimes\tau_0\,,\quad\gamma_{3}^{\,2}=\gamma_{5}^{\,2}=\mathds{1}_{4}  
\ee
The matrix $\gamma_{35}\equiv\mathrm{i}\gamma_{3}\gamma_{5}$ on the other hand commutes with $\gamma_{\mu}$ and anticommutes with $\gamma_{3}$ and $\gamma_{5}$.
The action of the Gross-Neveu model is invariant under global U($\Nf$)
transformations of the fermion fields. This also implies invariance under
$\text{U}(1)^{\otimes \Nf}$ transformations, i.e., the associated U($1$)-charge is conserved
in each flavor-sector separately. The matrix $\gamma_{35}$ further realizes a U$^{35}$($1$) symmetry in each flavor sector:
\be
\label{eq:vector_trafo}
\barspinor \mapsto \barspinor\mathrm{e}^{-\mathrm{i}\varphi\gamma_{35}},\,\quad\spinor \mapsto\mathrm{e}^{\mathrm{i}\varphi\gamma_{35}}\spinor\,.
\ee
{In addition to these two symmetries, the
model is symmetric under discrete $\mathds{Z}_{2}^{5}=\{\mathds{1}_{4},\gamma_{5}\}$ chiral transformations {acting on all flavors
simultaneously}: 
\be
\label{eq:chiral_trafo}
\bar\psi \mapsto -\bar\psi\gamma_{5}\,,\quad
\psi \mapsto \gamma_{5}\psi\,
\ee
(A similar symmetry transformation involving $\gamma_3$ can be understood as a
combination of $\mathds{Z}_{2}^{5}$ and U$^{35}$($1$) transformations.)}

In continuous dimensions $2<d<4$, we assume that a suitable analytic continuation for the
Dirac structure exists, such that traces over the algebraic structure yield
analytic functions in $d$ and $d_\gamma$.  The chiral symmetry of the model
can be associated with a $\mathds{Z}_2$ symmetry for the order parameter. As
we shall discuss below, the infrared regime of the theory is governed by
dynamical chiral symmetry breaking, provided the only parameter of the model,
namely $\bar{g}$, is adjusted accordingly.

\section{Asymptotic safety and RG flow equation }
\label{sec:AS}

For a self-contained presentation, let us briefly summarize the essentials of
Weinberg's asymptotic safety scenario which is based on the underlying
general structure of the renormalization group (RG). In the space of
parameters and couplings $g_i$, the RG provides a vector field $\boldsymbol
\beta$, summarizing the RG $\beta$ functions for these couplings $(\boldsymbol\beta)_i
= \beta_{g_i}(g_1,g_2,\dots)\equiv \pat g_i$. As the full content of a quantum
system can be parameterized in terms of generating functionals for correlation
functions, we can more generally study the RG behavior of a generating
functional. Introducing an IR-regulated effective average action $\Gamma_k$,
the RG flow of this action is determined by the Wetterich equation \cite{Wetterich:1992yh}
\begin{equation}\label{flowequation}
	\partial_t\Gamma_k[\Phi]
        =\frac{1}{2}\mathrm{STr}\{[\Gamma^{(2)}_k[\Phi]+R_k]^{-1}(\partial_tR_k)\},
        \;\, \pat=k\frac{d}{dk}  	.
\end{equation}
Here, $\Gamma^{(2)}_k$ is the second functional derivative with respect to the
field $\Phi$, representing a collective field variable for all bosonic or
fermionic degrees of freedom. The function $R_k$ denotes a momentum-dependent
regulator that suppresses IR modes below a momentum scale $k$.  The solution
to the Wetterich equation provides for an RG trajectory in the space of all
action functionals, also known as {\em theory space}, interpolating between
the bare action $S_\Lambda$ to be quantized $\Gamma_{k\to\Lambda}\to
S_\Lambda$ and the full quantum effective action $\Gamma=\Gamma_{k\to 0}$,
being the generating functional of 1PI correlation functions; for reviews, see
\cite{Litim:1998nf,Berges:2000ew,Rosten:2010vm}.

For a fundamental quantum field theory, the RG trajectory needs to be
extendable over all scales which, in particular, requires that the UV cutoff
$\Lambda$ can be sent to infinity. This is, for instance, possible if the
trajectory approaches a point in theory space which is a fixed point under the
RG transformations, i.e., remains invariant under the variation of
$k$. Parameterizing the effective average action $\Gamma_k$ by a
possibly infinite set of generalized dimensionless couplings $g_i$, the
Wetterich equation provides us with the flow of these couplings $\pat g_i=
\beta_{g_i}(g_1,g_2,\dots) $. A fixed point $g_{i,\ast}$ is defined by
\begin{equation}
 \beta_i(g_{1,{\ast}},g_{2,{\ast}},...)=0\ \forall \ i\,.
\end{equation}
The fixed point is non-Gau\ss ian, if at least one fixed-point coupling is
nonzero $g_{j,\ast}\neq 0$. If the RG trajectory hits a fixed point in the
UV, the UV cutoff can safely be taken to infinity and the system approaches a
conformally invariant state for $k\to\Lambda\to\infty$. 

For the theory to be predictive, the number of physical parameters required
for specifying the RG trajectory needs to be finite. Considering the
linearized flow in the fixed-point regime,
\begin{equation}
 \partial_t g_i = B_i{}^j (g_{j,\ast}-g_j)+\dots, \quad B_i{}^j =\frac{\partial
   \beta_{g_i}}{\partial g_j} \Big|_{g=g_\ast},\label{eq:lin}
\end{equation}
we encounter the stability matrix $B_i{}^j$, which we diagonalize
\begin{equation}
B_i{}^j\, V_j^I =-\Theta^I V_i^I,
\end{equation}
in terms of right-eigenvectors $V_i^I$, enumerated by the index $I$.
The resulting {\em critical exponents} $\Theta^I$ now provide for a classification
of physical parameters. The solution of the coupling flow in the linearized fixed-point
regime is given by
\begin{equation}
g_i=g_{i,\ast} + \sum_I C^I\, V_i^I \, \left(\frac{k_0}{k} \right)^{\Theta^I},
\end{equation}
where the integration constants $C^I$ define the initial conditions at a
reference scale $k_0$. All eigendirections with $\Theta^I<0$ are suppressed
towards the IR and thus are {\em irrelevant}. All {\em relevant} directions
with $\Theta^I>0$ increase towards the IR and thus determine the macroscopic
physics. For the {\em marginal} directions $\Theta^I=0$, higher-order terms in
the expansion about the fixed point have to be regarded. Hence, the number of
relevant and marginally-relevant directions determines the total number of
physical parameters to be fixed. The theory is predictive if this number is
finite.

For the flow towards the IR, the linearized fixed-point flow \Eqref{eq:lin} is
generally not sufficient and the full nonlinear $\beta$ functions have to be
taken into account. Even the parameterization of the effective action in terms
of the same degrees of freedom in the UV and IR might be inappropriate, for
instance, if bound states or condensates appear in the IR. This is precisely
the case in fermionic models beyond criticality; below, we discuss asymptotic
safety in the Gross-Neveu model therefore from both viewpoints, on the one
hand in terms of microscopic fermionic degrees of freedom and on the other
hand in terms of a mixed fermionic-bosonic description. 

\section{Fermionic fixed-point structure}\label{sec:fermion}
\label{sec:fermionic}

We begin with a discussion of  the  fixed-point structure of the Gross-Neveu
model as it becomes apparent already in a very elementary approximation within
a purely fermionic description. Let us consider only a point-like four-fermion
interaction, such that our ansatz for the effective action reads
\be
\Gamma_k[\psibar,\psi]=
\int_{x}\left\{Z_{\psi}\psibar\mathrm{i}\fslash{\partial}\,\psi+
  \frac{\bar{g}}{2 \Nf}(\psibar\psi)^2\right\}\,, 
\label{eq:eff_action_fermion}
\ee
where we allowed for a wave-function renormalization $Z_{\psi}$, and both
$Z_\psi$ and $\bar g$ are considered to be scale-dependent, i.e., a function
of $k$. This simple ansatz can be viewed as a derivative expansion of the
effective action, with the leading-order defined by $Z_{\psi}=$const. This
expansion can, in fact, be associated with a potentially small expansion
parameter in terms of the anomalous dimension $\eta_{\psi}=-\partial_t \ln
Z_{\psi}$.  Consequently, a running wave-function renormalization corresponds
to a next-to-leading order derivative expansion. Aside from derivatives,
further fermion channels and higher-order interactions compatible with the
symmetries can be taken into account. We come back to the role of such
interactions below.

Inserting the ansatz~\eqref{eq:eff_action_fermion} into the flow
equation~\eqref{flowequation}, the  flow of the
dimensionless renormalized four-fermion coupling~$g$,
{
\be
g =Z_{\psi}^{-2} k^{2-d} \bar{g}
\ee
}
is given by
\be
\beta_{g}\equiv\partial _t g = (d - 2 + 2\eta_{\psi})g - 4 d_{\gamma} v_d\,l_1^{F}(0)\, g^2\,,
\label{eq:fourfermionflow}
\ee
where $v_d^{-1}=2^{d+1}\pi^{d/2}\Gamma(d/2)$ and $\eta_{\psi}\sim {\mathcal
  O}(g)$. Here, we projected the full flow equation straightforwardly onto the
pointlike limit of the Gross-Neveu coupling; contributions from further
(possibly fluctuation-induced) interaction channels as well as dependencies on
the Fierz basis \cite{Jaeckel:2002rm,Jaeckel:2003uz,Gies:2009da} have been ignored for the {sake} of
simplicity. The constant $l_1^{F}(0)$ depends on the choice of the regulator
and parameterizes the regulator scheme dependence of the RG flow. For
instance, for a linear regulator of the form~\cite{Litim:2000ci,Litim:2001up,Litim:2001fd}
\be
R_k^{\psi}=Z_{\psi}\fslash{p}\,r_{\psi}(p^2/k^2)\,,\;
r_{\psi}(x)=\left(\frac{1}{\sqrt{x}}\!-\! 1\right)\Theta(1\!-\!x)\,,\label{eq:fermReg} 
\ee
where $p^2 = p_0^2 + \dots + p_d^2$, we have $l_1^{F}(0)=2/d$. Alternatively,
for a sharp cutoff, we find $l_1^{F}(0)=1$.  

\begin{figure}
\includegraphics[scale=0.7]{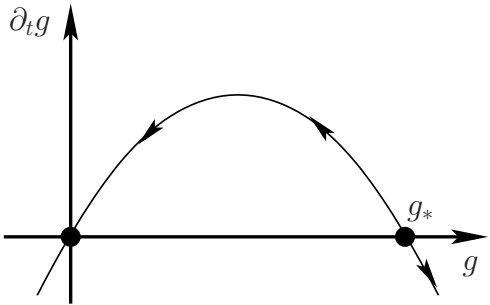}
\caption{Sketch of the $\beta$ function of the four-fermion coupling. For $g
  \geq g_{\ast}$ the infrared regime of the Gross-Neveu model is governed by
  chiral symmetry breaking (seemingly diverging fermionic interaction
  $g\to\infty$. For $g < g_{\ast}$ the model becomes a trivial (non-interacting)
  theory $g\to 0$ in the infrared.
 \label{fig:parabola}
}
\end{figure}

Apart from the Gau\ss ian fixed point we find a second non-trivial fixed point
for the coupling $g$ which is implicitly given by
\be
g_\ast=\frac{d - 2 + 2\,\eta_{\psi}(g_\ast)}{4 d_{\gamma}v_d\, l_1^{F}(0)}\,.
\ee
At leading order of our derivative expansion we have \mbox{$\eta_{\psi}\equiv 0$} 
and thus {
\be
g_\ast=\frac{d(d - 2)}{8 d_{\gamma}v_d}\stackrel{(d=3)}{=}\frac{3\pi^2}{4} \nn
\ee 
for the linear regulator and 
\be
g_\ast=\frac{d-2}{4 d_\gamma v_d}\stackrel{(d=3)}{=}\frac{\pi^2}{2}\nn
\ee
for the sharp cutoff.} The regulator dependence of the fixed-point value exemplifies the
non-universality of this quantity. Nevertheless, the existence of the fixed
point is a universal statement, as the regulator-dependent constant $l_1^F(0)$
is a positive number for any regulator. Moreover, the value of the non-Gau\ss
ian fixed point does not depend on $\Nf$. In Fig.~\ref{fig:parabola}, we show
a sketch of $\beta_{g}=\partial _t g$. The arrows indicate the direction of
the flow towards the infrared. The theory becomes trivial (non-interacting) in
the infrared regime for initial values $g_{\Lambda}< g_\ast$. Choosing
$g_{\Lambda} > g_\ast$, the four-fermion coupling increases rapidly
towards the infrared and diverges eventually. The divergence of $g$ at a
finite RG scale actually is an artifact of the over-simplistic fermionic
truncation, {but can be associated with the onset of chiral symmetry
breaking and the formation of a fermion condensate. {This becomes more 
obvious in the bosonic formulation, see below.}
In fact, the scale for a given IR observable $\mathcal O$ 
is set by the scale $k_{\rm cr}$ at which $1/g \to 0$: 
\be
{\mathcal O} \sim k_{\rm cr}^{d_{\mathcal O}}\,,
\ee
where $d_{\mathcal O}$ is the canonical mass dimension of the observable 
$\mathcal O$. For $g_{\Lambda} < g_{\ast}$, {the coupling never diverges
  but approaches zero}  
and thus ${\mathcal O}\equiv 0$. For $g_{\Lambda} > g_{\ast}$, 
 we find
\be
k_{\rm cr}=\Lambda \left(\frac{\left| g_{\Lambda} - g_{\ast}
\right|}{g_{\Lambda}}\right)^{\frac{1}{|\Theta|}},\label{eq:scaling} 
\ee
where the critical exponent $\Theta$ is given by
\begin{equation}
  \Theta=- \frac{\partial \beta_g(g=g_\ast)}{\partial g} =
  d-2+2\eta_{\psi,\ast}-2g_{\ast}\frac{\partial \eta_{\psi}}{\partial g}\Big|_{g_{\ast}}. 
\label{eq:fermTheta}
\end{equation}
Here, $\eta_{\psi,\ast}$ denotes the value of the fermionic anomalous dimension at the fixed point,
$\eta_{\psi,\ast}=\eta_\psi (g=g_\ast)$. Relation~\eqref{eq:scaling} determines how a given  
IR observable scales when $g_{\Lambda}$ is varied.
}
As the initial condition $g_\Lambda$
being larger or smaller than $g_\ast$ distinguishes between two different
phases in the long-range limit, the fixed point $g_\ast$ can be viewed as a
quantum critical point which divides the model into two physically different regimes.

For our main subject -- asymptotic safety --, we ignore the infrared physics
in the following and concentrate on the UV properties induced by the fixed
point. {In our simple fermionic truncation with only one coupling, the
stability matrix boils down to a number, already represented by the critical
exponent of the coupling direction in this one-dimensional theory space, 
see Eq.~\eqref{eq:fermTheta}.} At leading-order
derivative expansion, where $\eta_{\psi}=0$, the critical exponent is positive
for all $d>2$, such that the Gross-Neveu coupling corresponds to an RG
relevant coupling being attracted by the non-Gau\ss ian fixed point  towards
the UV. In this simple truncation, this suggests that the Gross-Neveu model can
be renormalized and extended as a fundamental theory over all scales on RG
trajectories that emanate from the non-Gau\ss ian fixed point. As there is only
one relevant direction, only one physical parameter has to be fixed (say the
value of the coupling at a UV scale, $g_\Lambda$) in order to predict all
physical quantities in the long-range limit. 

It is instructive, to compare these conclusions within the asymptotic safety
language with standard perturbation theory near the Gau\ss ian fixed point
$g_{\ast,\text{Gau\ss}}=0$. The corresponding critical exponent is
\begin{equation}
\Theta_{\text{Gau\ss}}= -  \frac{\partial \beta_g(g=g_{\ast,\text{Gau\ss})})}{\partial g} =
  2-d, \label{eq:Gaussexp}
\end{equation}
in agreement with (minus) the naive power-counting dimension of the coupling.
At leading-order derivative expansion $\eta_{\psi}=0$ and for $d>2$, the
critical exponent is negative and the fixed point thus infrared attractive. A
UV limit $\Lambda\to \infty$ can only be taken if the RG trajectory emanates
from the fixed point, but then the theory would be noninteracting on all
scales and therefore ``trivial''. Within perturbation theory, the conclusion
is that the Gross-Neveu model is perturbatively non-renormalizable. 
{Note that} this conclusion remains unchanged also if the anomalous dimension 
is taken into account: within perturbation theory, $\eta_\psi= \mathcal{O}(g)$
(actually, accidentally $\mathcal{O}(g^2)$), such that $\eta_\psi=0$ at the
Gau\ss ian fixed point, implying that standard power-counting can only be
modified logarithmically. 

Let us conclude with a word of caution on the derivative expansion in the
fermionic truncation: in this simple approximation, the fixed-point seems to
exist with similar properties in any dimension $d>2$, in particular also in
$d=4$ and beyond. This conclusion will change, once composite bosonic degrees
of freedom are taken into account. Fluctuations of the latter which are formed
by fermionic interactions will remove the fixed point in the Gross-Neveu model
for $d\geq 4$ such that no asymptotic safety scenario appears to exist for
$d\geq4$ in this model. In the fermionic language, the bosonic degrees of
freedom correspond to specific nonlocal interactions or momentum-structures in
the fermionic vertices. These are not properly resolved in a derivative
expansion. As $d=4$ is a marginal case, the conclusions for fermionic theories
in $d=4$ may depend on the details of the interaction and the algebraic
structure of a given model; for instance, an asymptotic safety scenario in a
standard-model-inspired $SU(\Nc)\times U(1)$ model has been discussed in
\cite{Gies:2003dp}. 

A comparison with the asymptotic safety scenario for quantum gravity is also
instructive: here, the upper critical dimension is $d=2$ as in the fermionic models,
and the non-Gau\ss ian fixed point exist in simple truncations based on
derivative expansions in all $d>2$
\cite{Litim:2003vp,Fischer:2006at}. It is tempting to
speculate whether strong-coupling phenomena such as bound-state formation may
destabilize the fixed point above another so far unknown critical dimension. A similar
phenomenon has been observed in extra dimensional Yang-Mills theories, where a
non-Gau\ss ian fixed point exists in $4+\epsilon$ dimensions but is
nonperturbatively destabilized for $\epsilon \gtrsim \mathcal{O}(1)$
\cite{Gies:2003ic}. 

\section{Partially Bosonized Gross-Neveu model}\label{sec:bos}

In this section, we study the (UV) fixed-point structure of the Gross-Neveu
model by employing a (partially) bosonized version of the model. We relate our
findings to the purely fermionic description, study analytically the
large-$\Nf$ limit and show how corrections beyond the mean-field approximation
can be systematically taken into account. Formulations of the Gross-Neveu
model using partial bosonization are used for many aspects of the Gross-Neveu
model such as the phase structure at zero and finite temperature and
density \cite{Hands:1995jq,Thies:2003kk,Castorina:2003kq}.

Spontaneous symmetry-breaking and the formation of a fermion condensate in the
Gross-Neveu model can conveniently be studied by introducing an auxiliary
field $\sigma$ into the functional integral. Formally, we introduce this
auxiliary field by multiplying the path-integral by a suitable Gau\ss ian
factor. This is known as a Hubbard-Stratonovich
transformation~\cite{Stratonovich,Hubbard:1959ub}. The partially bosonized
(PB) action of the Gross-Neveu model then reads 
\be S_{\rm
  PB}[\psibar,\psi,\sigma]=
\int_{x}\left\{\frac{\Nf}{2}\bar{m}^2\sigma^2
+\psibar\left(\mathrm{i}\fslash{\partial} 
    + \mathrm{i}\bar{h}\sigma\right)\psi\label{eq:PBaction} \right\}, 
\ee
and the functional integral now includes integration measures for both
fermionic and bosonic fields. As the auxiliary bosonic field only occurs
quadratically in the action, it can be integrated out again; on the level of
the action, this corresponds to replacing the bosonic field by its equation of
motion, $\Nf\,\bar{m}^2 \sigma = i \bar{h} \psibar\psi$. The action then
reduces again to the Gross-Neveu action upon identifying $ \bar{g} =
\frac{\bar{h}^2}{\bar{m}^2}$. From the viewpoint of the Hubbard-Stratonovich
transformation, the Yukawa coupling $\bar{h}$ is redundant as it can be scaled
into the sigma field. Only the ratio $\bar{h}^2/\bar{m}^2$ has a physical
meaning. In our formulation, the Yukawa coupling $\bar h$ is implicitly
understood to carry mass dimension $[\bar{h}]=(4-d)/2$ in order to deal with
an auxiliary field with canonical mass dimension $[\sigma]=(d-2)/2$. Moreover,
the $\sigma$ field has been scaled such that the first and second term
in~\eqref{eq:PBaction} are of the same order in $\Nf$. Under a discrete chiral
transformation, see Eq.~\eqref{eq:chiral_trafo}, the $\sigma$-field transforms
as $\sigma \mapsto -\sigma$. From a phenomenological point of view $\sigma$
can be considered as a bound state of fermions, $\sigma \sim
\psibar\psi$. Thus, the vacuum expectation value of $\sigma$ is a proper order
parameter for chiral symmetry breaking, as it is the case in the purely
fermionic formulation of the model.

\subsection{Mean-field analysis}

Before we analyze the partially bosonized version of the Gross-Neveu model by
means of the functional RG, let us start with a simple mean-field study,
corresponding to the large-$\Nf$ limit, in order to rediscover aspects of the
fermionic language of the preceding section in this standard textbook
language. Due to the Hubbard-Stratonovich transformation, the $\Nf$ fermion
flavors enter only bilinearly and can be integrated out from the corresponding
functional integral. This yields a purely bosonic effective theory for the
Gross-Neveu model:
\be
S_{\rm B}[\sigma]= \int_{x}\,\frac{\Nf}{2}\bar{m}^2\sigma^2 - \Nf
\mathrm{Tr}\ln\left[\mathrm{i}\fslash{\partial} +
  \mathrm{i}\bar{h}\sigma\right], 
\ee
where ${\rm Tr}$ denotes a functional trace. Since $\sigma$ depends on the
space-time coordinates, the action $S_{\rm B}$ is highly non-local and
therefore in general difficult to study. The ground state $\sigma_0$ can be
obtained from the variational principle, i.e., the gap equation
\be
\frac{\delta}{\delta\sigma(x)}S_{\rm B}[\sigma]\Bigg|_{\sigma_0(x)}\stackrel{!}=0.\label{eq:Baction}
\ee
As we are interested in the UV properties of the model, we assume that the
ground state is homogeneous. (In $d=2$, inhomogeneous condensates have been
identified in some parts of the phase diagram at finite temperature and large values of the chemical 
potential~\cite{Thies:2003kk}; the
status for higher-dimensional fermionic models is subject to ongoing work,  see e.~g.~\cite{Nickel:2009wj,Kojo:2009ha})
Such a homogeneous ground state $\sigma_0=$const. is then implicitly given by the solution of the following equation:
\be\label{eq:meanfield}
\sigma_0=4\frac{\bar{h}^2}{\bar{m}^2}\int_{p}\frac{\sigma_0}{p^2 + \bar{h}^2 \sigma_0^2}\,,
\ee
with $\int_{p}=\int\frac{d^d p}{(2\pi)^d}$. Apparently this equation has a
trivial solution, $\sigma_0=0$. Moreover, non-trivial solutions for $\sigma_0$
can be found for suitably adjusted values of the four-fermion coupling
$\bar{g} =\bar{h}^2/\bar{m}^2$. Since in $d=3$ ($d=4$) the right-hand side
of \Eqref{eq:meanfield} is linearly (quadratically) divergent, we impose a
sharp UV cutoff $\Lambda\gg m_{\rm f}=\bar{h}^2\sigma_0^2$.  For $d=3$, we find
{
\begin{equation}\label{eq:mfmass}
m_{\rm f} = \frac{2}{\pi}\left( \Lambda - \frac{\pi^2}{2}\frac{\bar{m}^2}{\bar{h}^2}\right)\,.
\end{equation}
}
Thus, the fermions acquire a non-zero mass due to the spontaneous breakdown of
chiral symmetry, provided we choose $\bar{m}^2/\bar{h}^2 < 2\Lambda/\pi^2$. In
terms of the four-fermion coupling, we can read off a critical value for the
dimensionless coupling {$g=\Lambda \bar{g}=\Lambda \bar{h}^2/\bar{m}^2$} 
above which the IR physics is governed by spontaneous symmetry breaking:
{
\begin{equation}
g_{\rm cr}=\frac{\pi^2}{2}\,.\label{eq:mfcrit}
\end{equation}
} This critical value $g_{\rm cr}$ can be identified with the sharp-cutoff
value of the non-trivial fixed point $g_\ast$ which we found in our study of
the fermionic fixed-point structure. The role of the critical value as a fixed
point becomes obvious from the fact that if $g=g_{\rm cr}$ the theory is
interacting but remains massless and ungapped on all scales. From the
viewpoint of the partially bosonized theory, we find that the Yukawa coupling
and the boson mass are not independent parameters. For a fixed ratio
$\bar{h}^2/\bar{m}^2 > g_{\rm cr}$ the IR physics remains unchanged. Thus, the
purely bosonic description of the theory {in this approximation} depends
only on a single parameter as in the fermionic formulation.

\subsection{RG flow of the partially bosonized theory}

Let us now discuss the fixed-point structure of the partially bosonized
theory. A partially bosonized description of the theory is appealing from a
field-theoretical point as it also forms the basis for the expansion in
$1/\Nf$ for a large number of flavors. In addition, it allows us to
systematically resolve parts of the momentum dependence of the vertices by
means of a derivative expansion. As we shall see below, these two expansion
schemes are not identical and should therefore not be confused with each
other.
For our study we employ the following ansatz for the effective action:
\be
\Gamma[\psibar,\psi,\sigma]&=&\int_{x}\left\{\frac{{\Nf}}{2}Z_{\sigma}(\partial_{\mu}\sigma)^2+
\psibar\left(Z_{\psi}\mathrm{i}\fslash{\partial} + \mathrm{i}\bar{h}\sigma\right)\psi\right.\nonumber\\
&&\left. \qquad\qquad +\Nf\, U(\sigma^2)
\label{eq:PBgamma}
\right\}\,,
\ee
where we allow all couplings and wave function renormalizations
$Z_{\sigma,\psi}$ to be scale dependent. The kinetic term of the boson field
adds a new aspect: on the one hand, it goes beyond the local approximation of
simple mean-field theory; in terms of the fermionic language, this kinetic
term corresponds to a specific momentum dependence in the scalar $s$ channel
of the four-fermion coupling on the other hand.  As we shall see below, this
term and the associated wave function renormalization receive contributions to
leading order in the large-$\Nf$ approximation {(in order to simplify the
  large-$\Nf$ counting of orders, the purely bosonic sector is multiplied by
  an $\Nf$ factor in \Eqref{eq:PBgamma} similar to \Eqref{eq:PBaction})}. The 
large-$\Nf$ flow corresponds to the choice
\be
Z_{\sigma}\big|_{k\to\Lambda}\to 0\,,\qquad \partial_t Z_{\sigma}\neq 0\,,\nonumber\\
Z_{\psi}\big|_{k\to\Lambda}\equiv 1\,,\qquad \partial_t Z_{\psi}\equiv 0\,.
\ee
This exemplifies the difference between large-$\Nf$ and derivative expansion,
as with this choice we include next-to-leading order corrections in terms of a
derivative expansion in the bosonic sector but treat the fermionic sector in
the leading-order approximation. 

From our ansatz~\eqref{eq:PBgamma} we see that a {nonzero homogeneous
expectation value for $\mathrm{i}\bar{h}{\sigma}$ plays the role of a mass term for
the fermions. By expanding the effective action about the homogeneous
background ${\sigma}$} we anticipate that condensation occurs only in the
homogeneous channel.  The bosonized Yukawa model is fixed to the fermionic
Gross-Neveu model by a suitable choice of initial conditions for $\Gamma$ at
the UV scale $\Lambda$. This correspondence is established by the choice
\be
Z_{\sigma}\big|_{k\to\Lambda} \ll 1\,, \quad Z_{\psi}\big|_{k\to\Lambda} \to 1\,,\quad 
U_{\Lambda}=\frac{1}{2}\bar{m}^{2}_{\Lambda}\,. 
\ee
Thus, the renormalized boson mass $m=\bar{m}/\sqrt{Z_{\sigma}}$ at the UV cutoff
$\Lambda$ becomes much larger than $\Lambda$ and renders the boson propagator
essentially momentum independent. This compositeness condition for the bosonic
formulation can be considered as a locality condition at the UV scale for
the four-fermion coupling $g$ in the purely fermionic formulation of the
model.

Since we are interested in the UV fixed-point structure of {the}
partially bosonized (chirally-symmetric) Gross-Neveu model, {we
  anticipate that a possible non-Gau\ss ian fixed point occurs in the
  symmetric regime}. Therefore, we only need to study the RG flow in the
symmetric regime with vanishing vacuum expectation value for the
$\sigma$ field. We then find the following flow equation for the dimensionless
effective potential:
\be
\partial_{t}u  &=&   -d u + (d-2+\eta_{\sigma})u^{\prime}\rho -2d_{\gamma}v_d\, l_0^{(F)d}(2 h^2\rho;\eta_{\psi}) \nn\\ 
&& \qquad+ \frac{1}{\Nf} 2 v_d \,l_0^{d}(u^{\prime}+2\,\rho
u^{\prime\prime};\eta_{\sigma})\,, 
\label{eq:patu}
\ee
where 
\be
u(\rho)= k^{-d}U(\sigma)\,,\qquad \rho=\frac{1}{2}Z_{\sigma}k^{2-d}\sigma^{2},\label{eq:u_flow}
\ee
and the dimensionless renormalized Yukawa coupling is given by
\be
 h^{2}=Z_{\sigma}^{-1}\wavpsi^{-2} k^{d-4}\bar{h}^{2}\,.
\ee
{
The flow equation for the potential can be solved either directly or by an 
expansion around $\rho=0$, 
\be
u(\rho)=\sum_{n=0}^{\infty}\frac{\lambda_{2n}}{n!}\rho^n\,.
\ee
} In this paper, we shall mainly focus on the latter one which allows us to
directly project {onto} the (dimensionless) bosonic mass parameter $\lambda_2$ and
the higher-order bosonic couplings $\lambda_{2n}$:
\be
\lambda_2=\frac{\bar{m}^2}{Z_{\sigma}k^2}\,,\qquad \lambda_{2n}=Z_{\sigma}^{-n} k^{(n-1)d-2n} U^{(n)}\big|_{\sigma =0}\,.
\ee
{The flow equations for the Yukawa coupling as well as the anomalous
  dimensions $\eta_{\sigma}=-\partial_t \ln Z_{\sigma}$ and
  \mbox{$\eta_{\psi}=-\partial_t \ln Z_{\psi}$}} read\footnote{{These flow
    equations agree with those derived in \cite{Hofling:2002hj,Gies:2009hq} in
    the symmetric phase, upon a rescaling of the wave function renormalization
    $Z_\sigma$ by a factor of $\Nf$, cf. \Eqref{eq:PBgamma}.}}
{
\be
\partial_{t}h^2  &=&  (d-4 + 2\eta_{\psi}+\eta_{\sigma})h^2  \nn\\
&& \qquad +\frac{1}{\Nf}8 v_d\,h^4 \, l_{1,1}^{(FB)d}(0,\lambda_2;\eta_{\psi},\eta_\sigma)\,, \label{eq:hflow}\\ 
\eta_{\sigma}  &=&  8
\frac{d_{\gamma}v_d}{d}h^2\,m_{4}^{(F)d}(0;\eta_{\psi})\,, \label{eq:etasigma}\\ 
\eta_{\psi}  &=& \frac{1}{\Nf} 8 \frac{v_{d}}{d}h^{2}m_{1,2}^{(FB)d}(0,\lambda_2;\eta_{\psi},\eta_{\sigma})\,,\label{eq:yuk_flow}
\ee
}
The threshold functions $l$ and $m$ in Eqs.~\eqref{eq:u_flow} and
\eqref{eq:hflow}-\eqref{eq:yuk_flow} depend on the details of the
regulator. {For practical computations, we use an optimized 
regulator~\cite{Litim:2000ci,Litim:2001up,Litim:2001fd} 
for the fermionic fields~\eqref{eq:fermReg} and the bosonic fields:
\be 
R_k=Z_{\sigma}r(p^2/k^2)\,,\quad r(x)=\left(\frac{1}{x}-1\right)\Theta(1-x)\,.
\label{eq:bosReg}
\ee
}
{The corresponding threshold functions are listed in App.~\ref{App:tf}. 
These functions essentially describe the threshold behavior 
of regularized 1PI diagrams.}

\subsection{RG flow at large $\Nf$}

\begin{figure}[t]\center
\includegraphics[scale=0.75]{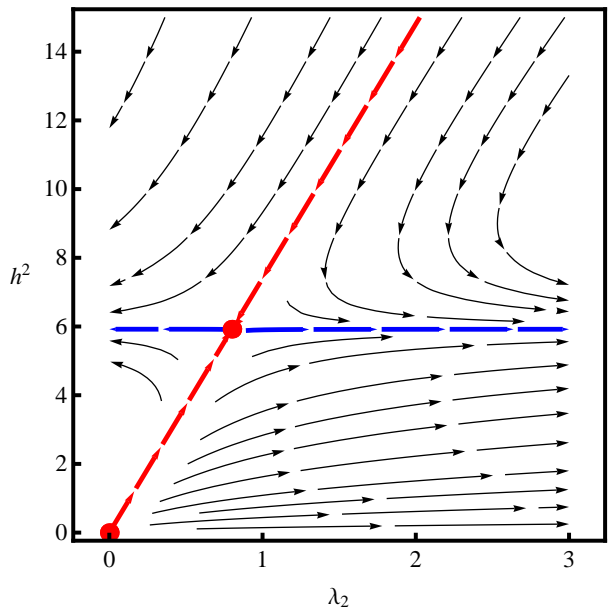}
\caption{\label{fig:loflow} Leading-order RG flow in the $1/\Nf$ expansion for
  the partially bosonized Gross-Neveu model in the $(h^2,\lambda_{2})$
  plane. The red line denotes the critical manifold {of points drawn into
    the fixed point towards the infrared}. The blue line {depicts the
    critical surface attracting the flow towards the IR.} {The critical
    surface contains all trajectories that emanate from the non-Gau\ss ian
    fixed point in the UV; its dimensionality equals the number of relevant
    directions and thus the number of physical parameters to be fixed}. The red dots denote
  the Yukawa Gau\ss ian fixed point and the non-Gau\ss ian fixed point,
  respectively. The arrows indicate the direction of flow towards the
  infrared.}
\end{figure}

As the nonperturbative renormalizability of the Gross-Neveu model beyond $d=2$
has been proved to all orders in the $1/\Nf$ expansion, it is worthwhile to
study asymptotic safety on the basis of the RG flow in the same limit. Here,
our set of flow equations for the partially bosonized Gross-Neveu model
reduces to
\be
\partial_{t}u  &=&   -d u + (d-2+\eta_{\sigma})u^{\prime}\rho \nn\\
&& \qquad -2d_{\gamma}v_d\, l_0^{(F)d}(2 h^2\rho;\eta_{\psi}), \\ 
\partial_{t}h^2  &=&  (d-4 + 2\eta_{\psi}+\eta_{\sigma})h^2, \\
\eta_{\sigma}  &=&  8 \frac{d_{\gamma}v_d}{d}h^2\,m_{4}^{(F)d}(0;\eta_{\psi})\,, \\
\eta_{\psi}  &=& 0\,.\label{eq:largenf_flow}
\ee
As the bosonic fluctuations carry no flavor number, we observe that 1PI
diagrams with at least one inner bosonic line decouple completely from the
large-$\Nf$ RG flow. As a consequence, $\eta_{\sigma}$ is non-vanishing in
leading order in the $1/\Nf$ expansion, whereas the fermionic anomalous
dimension is zero.\footnote{{It is worthwhile to emphasize, that the
    large-$\Nf$ counting is very different from scalar O($N$) models, where
    the the anomalous dimensions are zero to leading order at large $\Nf$
    \cite{ZinnJustin:2002ru}. Also the structure of the potential equation is
    very different, such that also the search for an exact solution requires a
    different strategy from that of O($N$) models \cite{Tetradis:1995br}, see
    below.}} The flow equations for the bosonic couplings are essentially
driven by the fermion loop,
\be
\partial_{t}\lambda_{2 n} &= & \, \bigl(n(d-2 + \eta_{\sigma}) - d\bigr)\lambda_{2 n}  \nn\\ 
&& \qquad- (-1)^{n} n!\,2^{n+2}\left(\frac{d_{\gamma}v_{d}}{d}\right)\left( h^{2}\right)^{n}.
\ee
Fixed point values $h_{\ast}^{2}$, $\eta_{\sigma,\ast}$, $\lambda_{2,\ast}$,
$\lambda_{4,\ast}$, $\lambda_{6,\ast}$, $\dots$ can be identified as the zeroes of 
the corresponding $\beta$ functions,
$\partial_{t}h^2\stackrel{!}=0\,,\; \partial_{t}\lambda_2\stackrel{!}=0, \dots$ 
Of course, the Gau\ss ian fixed point with all couplings vanishing solves
these fixed point equations.  As the RG flows for the bosonic couplings
decouple in the large-$\Nf$ limit, a non-trivial fixed point requires
$h_{\ast}\neq0$. This immediately requires
\be
\eta_{\sigma,\ast}=4-d. \label{eq:etaNf}
\ee 
Whereas this tight relation between the dimensionality and the bosonic
anomalous dimension here is an artifact of the large-$\Nf$ expansion, a
similar relation exists in gravity for the graviton anomalous dimension at the
fixed point as a consequence of background gauge invariance. {Similar sum
  rules are known for Yukawa theories with chiral symmetries
  \cite{Gies:2009da}. Such a sum rule for a corresponding fixed point is also
  responsible for the universality of the BCS-BEC crossover in the broad
  resonance limit of ultracold fermi gases \cite{Diehl:2007th}.}
In the present case, this fixes the value of the Yukawa fixed-point coupling,
\be
h_{\ast}^2=\left(\frac{d}{d_{\gamma}v_{d}}\right)\frac{(d-4)(d-2)}{(8-6 d)}.
\label{eq:hFP}
\ee
The fixed point is interacting for $2<d<4$ and merges with the Gau\ss ian
fixed point in $d=2$ and $d=4$. Also the fixed-point values for the bosonic mass
parameter and couplings can be given analytically
\be
\lambda_{2 n,\ast}\!=\! \frac{(-1)^{n}n!\,2^{n+2}}{(2n\! -\! d)}
\left(\frac{d}{d_{\gamma} v_{d}}\right)^{n-1}\left(\frac{(d\!-\! 4)(d\! -\!
    2)}{(8\! -\! 6 d)}\right)^{n}\!\! .
\ee
Thus, the fixed-point values for all bosonic vertices $\lambda_{2n}$ are
non-vanishing. In a purely fermionic formulation of the model, these higher
bosonic self-interactions correspond to higher (non-local) fermionic
self-interactions. In any case, we find that the UV fixed-point theory in
$2<d<4$ for $\Nf\to\infty$ is not identical to the action $S_{\rm PB}$ but
involves infinitely many operators. 

\begin{figure}[t]\center
\includegraphics[scale=0.75]{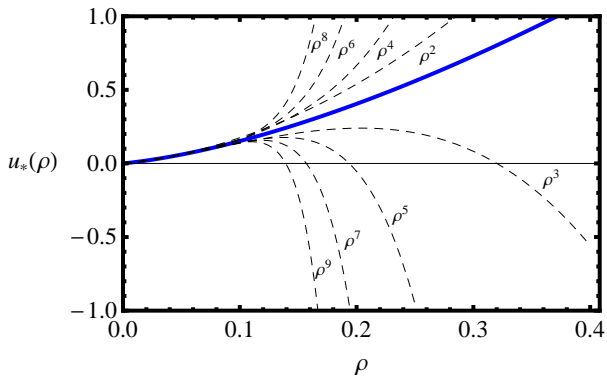}
\caption{\label{fig:fixpot} 
(Resummed) Fixed-point potential $u_{\ast}(\rho)$ in the large-$\Nf$ limit (blue/solid), 
see Eq.~\eqref{eq:fppot}, and low-order polynomial approximations thereof ranging from $\rho^{2}$ to $\rho^{9}$
for small values of $\rho$ (dashed lines).
}

\end{figure}

It turns out that the alternating series
for the full renormalized effective potential can be resummed and yields a representation
of the Gau\ss ian Hypergeometric function $_{2}F_{1}(a,b;c;z)$ ({see e.g. \cite{GR}}). In terms
of renormalized fields, the scale invariant fixed-point action then describes massless fermions coupled
to a scalar boson with a potential
\begin{align}
u_{\ast}(\rho)=&-\frac{2d-8}{3d-4}\rho\,\,\times\nn\\ 
&_{2}F_{1}\left(1-\frac{d}{2},1;2-\frac{d}{2};\frac{(d-4)(d-2)}{6d-8}\frac{d}{d_{\gamma}v_{d}}\rho\right).
\label{eq:fppot}
\end{align}
This potential has a large-field asymptotic behavior $\propto \rho^{\frac{3}{2}}$. The small-field region is
depicted in Fig.~\ref{fig:fixpot}.

The theory has predictive power, as the number of physical parameters is
determined by the number of RG relevant directions corresponding to the number
of positive critical exponents.
{At the} non-Gau\ss ian fixed point in the large-$\Nf$ limit, the stability matrix 
{assumes} a particularly simple form where only a single column and the main diagonal 
are non-vanishing,
\begin{equation}
B=
\begin{pmatrix}
b_{h^2,h^2} & 0 & \cdots & & \\
b_{h^2,\lambda_{2}} & b_{\lambda_{2},\lambda_{2}} & 0 & \cdots &  \\
b_{h^2,\lambda_{4}} & 0 & b_{\lambda_{4},\lambda_{4}} & 0 & \cdots & \\
\vdots & \vdots & 0 & \ddots & 
\end{pmatrix}.
\end{equation}
{Therefore} only the main diagonal of $B$ enters into the stability analysis around the non-Gau\ss ian fixed point.
The important non-vanishing entries $b_{h^{2},h^{2}}$ and
$b_{\lambda_{2n},\lambda_{2n}}$ {turn out to be completely universal and}
are given by
\begin{equation}
b_{h^{2},h^{2}}\equiv\frac{\partial\beta_{h^{2}}}{\partial h^2}\Big|_{g=g_\ast}=\eta_{\sigma,\ast}=4-d
\end{equation}
and
\begin{equation}
b_{\lambda_{2n},\lambda_{2n}}\equiv\frac{\partial\beta_{\lambda_{2n}}}{\partial\lambda_{2n}}\Big|_{g=g_\ast} =-d+n(d-2+\eta_{\sigma,\ast})=2n-d.
\end{equation}
The characteristic polynomial $\mathrm{det}(B+\Theta\mathds{1})$ of the matrix $B$ yielding the eigenvalues $-\Theta^{I}$ 
via its zeroes is then easily found to be 
\begin{equation}
(-\Theta-(4-d))\prod_{n=1}^{\infty}\left(-\Theta-(2n-d)\right).
\end{equation}
{All large-$\Nf$} critical exponents {are thus} given by {$d-4$
  and $d-2n$}. In $d=3$ this boils down to one positive critical exponent with
value $1$, i.e., one relevant RG direction, and infinitely many negative
{critical exponents $-1,-1,-3, -5, -7, \dots$}, corresponding to
irrelevant RG directions.
For the case of the Gau\ss ian Yukawa fixed point, the characteristic polynomial of the stability 
matrix $B$ is changed to  
\begin{equation}
(-\Theta-(d-4))\prod_{n=1}^{\infty}\left(-\Theta-(n(d-2)-d)\right).
\end{equation}
As expected, the critical exponents coincide with the mass dimension of the
Yukawa coupling and the bosonic couplings, {reproducing simple
  perturbative power counting}. In total, this yields three relevant RG
directions and one marginal RG direction.  {Note} that the Gau\ss ian
fixed point found in the purely fermionic flow in Sect.~\ref{sec:fermion}
translates into a diverging {dimensionless} renormalized boson mass
{$\lambda_{2}^{1/2}\sim g^{-1/2}$.}  

{Returning to the non-Gau\ss ian fixed point, we can, of course, make
  contact with the purely fermionic description and} deduce the fixed point 
  of the four-fermion coupling from the fixed point values of the Yukawa
  coupling and the bosonic mass parameter. We find {
\be
g_\ast=\frac{h^2_{\ast}}{\lambda_{2,\ast}}=\left(\frac{d}{d_{\gamma}v_{d}}\right)\frac{d-2}{8}
\stackrel{(d=3)}{=}\frac{3\pi^2}{4},
\ee
{for the linear regulator} which agrees with our findings in
Sect.~\ref{sec:fermion} (using the {same} regulator).}  { In fact, the
flow equation of the four-fermion interaction $g$ can be {reconstructed}
from the flow of $h^2/\lambda_2$,
\be
\partial_t \left(\frac{h^2}{\lambda_2}\right)\!=\!(d\!-\!2)\left(\frac{h^2}{\lambda_2}\right)
\!-\!\frac{8d_{\gamma}v_d}{d}\left(\frac{h^2}{\lambda_2}\right)^2
\!+\!{\mathcal O}\left(\frac{1}{\Nf}\right)\!.
\ee
Using the linear regulator in Eq.~\eqref{eq:fourfermionflow}, we observe that the flow equation
for $g$ and $h^2/\lambda_2$ are identical in the large-$\Nf$ limit. Recall that
$\eta_{\psi}=0$ in this limit.
}

Due to the equivalence of $g$
and $h^2/\lambda_2$, the quantum critical point found in the purely fermionic
formulation is also present in our study of the partially bosonized theory for
$\Nf\to\infty$, as it should be the case. 
As it can be seen from the scaling law \eqref{eq:scaling}, this quantum critical
point is associated with a vanishing boson mass, i.e., a diverging correlation
length, in the long-range limit.

Let us conclude our large-$\Nf$ analysis with a word of caution on the
  widely used so-called local potential approximation {(LPA)} in which
  the running of the wave-function renormalizations are neglected. If we
  {ignored} the running of the wave-function renormalization of the
  bosonic field {in the present case}, the model would
  {artificially} depend on more than one {physical} parameter.  To
  be more specific, {let us} consider the mass spectrum of the theory in
  the regime with broken chiral symmetry and assume that we have
  {already} fixed the mass of the fermions. Using the definition of the
  masses and the flow equations of the couplings, we find that the
{
(dimensionless) renormalized boson mass in the broken regime 
can be written in terms of the (renormalized)
  fermion mass $m_{\rm f}$}:
\be
m^2 =2\lambda_4 \rho_0
\sim Z_{\sigma}^{-1}\bar{h}^2(\bar{h}^2\bar{\rho}_0)\sim 
Z_{\sigma}^{-1}\bar{h}^2 m_{\rm f}^2\,,
\ee
where $\rho_0=(1/2)\sigma_0^2$.  Neglecting the running of $Z_{\sigma}$,
i.~e. $Z_{\sigma}={\rm const.}$ as is done in the LPA, we observe that
the boson mass does not depend on a single physical parameter, as it should
be, but on two parameters independently, namely the fermion mass and
the (bare) Yukawa coupling. By contrast, taking the running of
$Z_{\sigma}\sim \bar{h}^2$ into account, the value of {the} boson mass is fixed
solely in terms of the fermion mass, in agreement with our
  fixed-point analysis. 
{
While this argument might be altered in $d=4$ 
space-time dimensions where the Yukawa coupling is marginal, 
it is true for the Gross-Neveu model (as well as the Nambu-Jona-Lasinio model) 
in any dimension $d$ {in which} the flow equation for $Z_{\sigma}$ is non-vanishing even {at} 
leading order in an expansion in $1/\Nf$. Therefore the flow of $Z_{\sigma}$ has to be taken into account in 
a systematic and consistent expansion of the flow equations in powers of $1/\Nf$. To be
specific, the flow of the order parameter potential~\eqref{eq:patu} incorporates already
fluctuations at next-to-leading order in $1/\Nf$ due to the presence of the bosonic 
loop\footnote{
{
We stress that our parameter $\Nf$ plays the role of the number of colors $\Nc$ 
in QCD. The number of flavors $\Nf$ in, e.~g., 
QCD low-energy models {is} related to the number of involved
{mesonic} scalar fields $N=\Nf^2$. Thus,
a large-$\Nc$ expansion in QCD models corresponds to a large-$\Nf$ expansion in the
Gross-Neveu model. For an analysis of the role of corrections beyond the large-$\Nc$
approximation for the thermodynamics of QCD low-energy models we refer to Ref.~\cite{Braun:2009si}.
}
}. 
However, for a systematic and consistent study of the effects of corrections beyond 
the large-$\Nf$ expansion the flow of $Z_{\sigma}$, $Z_{\psi}$ as well as of the Yukawa 
coupling needs to be taken into account.
} 

\subsection{RG flow for general flavor number $\Nf$}

Beyond the limit of large $\Nf$, bosonic fluctuations begin to play a role. An
immediate consequence is that a new fixed point for $h=0$ arises for the
flow of the effective potential for $2<d<4$. This fixed point of the purely
bosonic theory is nothing but the Wilson-Fisher fixed point which describes
critical phenomena in the Ising universality class. 

The non-Gau\ss ian fixed point of the full Gross-Neveu system can now be
  understood as being sourced from the leading large-$\Nf$ terms discussed
  above and the bosonic fluctuations inducing a Wilson-Fisher fixed
  point. Depending on the value of $\Nf$ the non-Gau\ss ian Gross-Neveu fixed
  point interpolates between the large-$\Nf$ fixed point for $\Nf\to \infty$
  and the Wilson-Fisher fixed point in the formal limit of $\Nf\to0$. For the
  latter, the functional RG has already proven to be a useful quantitative
  tool for describing nonperturbative critical {phenomena, see 
e.~g. Refs.~\cite{Tetradis:1993ts,Berges:2000ew,Litim:2001hk,Litim:2002cf,Bervillier:2007rc,Benitez:2009xg,Litim:2010tt}.}

\begin{figure}
\includegraphics[scale=0.75]{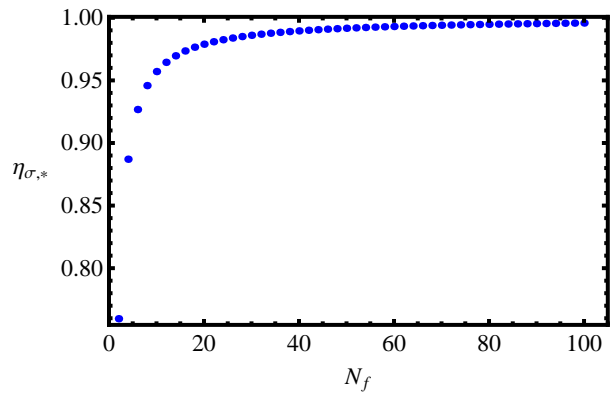}
\caption{\label{fig:etasig} Scalar anomalous dimension $\eta_\sigma$ for
  $\Nf=2,4,6,\dots,100$ in $d=3$.}
\end{figure}

\begin{figure}
\includegraphics[scale=0.75]{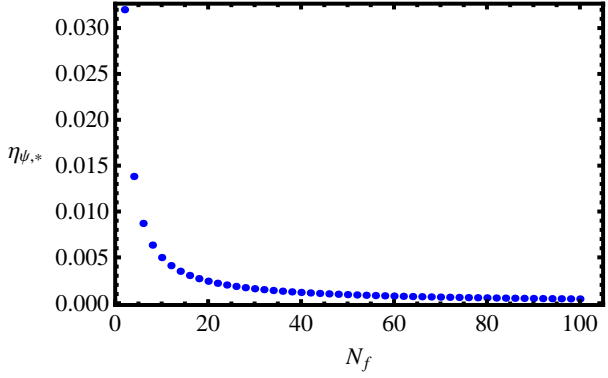}
\caption{\label{fig:etapsi} 
Scalar anomalous dimension $\eta_\psi$ for $\Nf=2,4,6,\dots,100$ in $d=3$.}
\end{figure}

Let us repeat the preceding large-$\Nf$ analysis, now using the full flow
equations at next-to-leading order derivative expansion,
i.e., Eqs.~\eqref{eq:patu}, \eqref{eq:hflow}, \eqref{eq:yuk_flow}, and
\eqref{eq:etasigma}. {For all quantities of interest, such as critical
  exponents and fixed point values of couplings, a solution of the potential
  flow in a polynomial expansion is sufficient. Confining our numerical
  studies to $d=3$, all figures are produced within a truncated expansion up
  to 22nd order in $\sigma$; quantitative results are derived from an expansion to the same order
  in $\sigma$. In the
  symmetric regime, a nontrivial fixed point in the Yukawa coupling requires
  the following inequality to be satisfied,
\begin{equation}
d-4 + 2\eta_{\psi,\ast}+\eta_{\sigma,\ast}<0, \quad \text{for}\,\,\Nf<\infty. \label{eq:ineq}
\end{equation}
{This is because} the second term of the Yukawa flow \Eqref{eq:hflow} is
strictly positive for admissible values of the anomalous dimensions
$\eta_\sigma,\eta_\psi \lesssim \mathcal{O}(1)$. For instance, in $d=3$, the
sum of the anomalous-dimension terms is always slightly smaller than 1, see
Figs. \ref{fig:etasig} and \ref{fig:etapsi}. The inequality becomes an
equality in the large-$\Nf$ limit, see \Eqref{eq:etaNf}.
}

\begin{figure}
\includegraphics[scale=0.75]{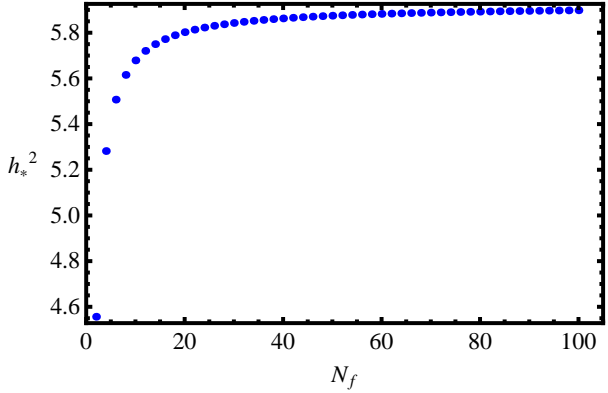}
\caption{\label{fig:Yuk} 
Yukawa fixed-point value over $\Nf=2,4,6,\dots,100$ in $d=3$.}
\end{figure}

The resulting fixed point values for the Yukawa coupling in $d=3$ are depicted in
  Fig. \ref{fig:Yuk}. For increasing $\Nf$, the fixed point quickly approaches its
  large-$\Nf$ limit \eqref{eq:hFP}, whereas it tends to zero for small $\Nf$
  leaving us with the pure Wilson-Fisher fixed point of a pure scalar
  model. As the latter is known to exhibit a fixed-point potential in the
  broken regime, i.e., with a non-vanishing expectation value of the scalar field $\sigma$, we expect
  such a fixed-point potential featuring a nontrivial minimum to occur for small $\Nf$. For all
  integer values of $\Nf\geq 1$ Dirac (four-component) fermions, we still observe
  fixed-point potentials in the symmetric regime, in agreement with
  \cite{Hofling:2002hj}. Nevertheless, the fixed point seems to occur in the
  broken regime for the
  model with one two-component fermion (corresponding to $\Nf=1/2$ in our
  language) \cite{Hofling:2002hj}. 

\begin{figure}
\includegraphics[scale=0.68]{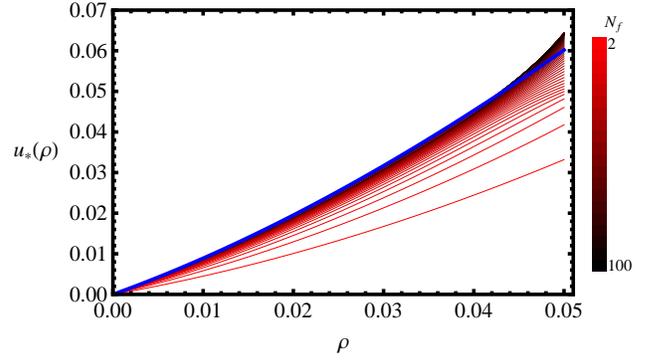}
\caption{\label{fig:potplot} Fixed-point potentials $u_{\ast}(\rho)$ for
  $\Nf=2,4,6,\dots,100$ in $d=3$. The
  large-$\Nf$ result is shown {as a thick/blue line. Incidentally, for
  $\Nf=0$ the potential approaches the Wilson-Fisher fixed-point potential
  with a nontrivial minimum {near $\rho\simeq 0.3$} (not shown).} }
\end{figure}

The fixed-point potential in $d=3$ for various values of $\Nf$ is
plotted in Fig.~\ref{fig:potplot} in a 22nd-order approximation. Also the 
potential converges rapidly to the large-$\Nf$ result for increasing values
of $\Nf$.

\begin{figure}

\includegraphics[scale=0.75]{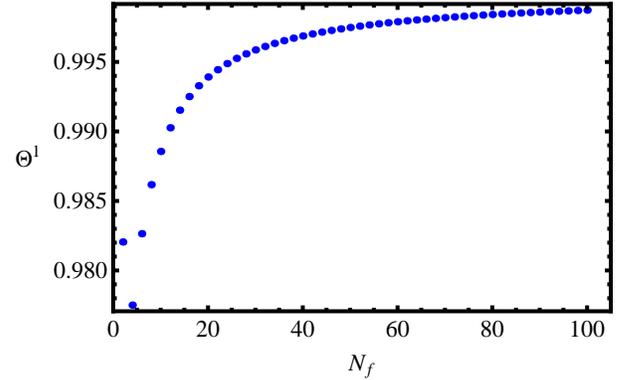}
\caption{\label{fig:theta1} 
Relevant critical exponent $\Theta^{1}$ for $\Nf=2,4,6,\dots,100$ in
$d=3$. {The non-monotonic behavior for small $\Nf$ is expected, since
  $\Theta^1$ has to approach $\Theta^1\simeq1.6$ for $\Nf=0$ corresponding to a
  correlation length exponent $\nu=1/\Theta^1\simeq0.63$ of the Ising
  universality class.  }
}
\end{figure}

\begin{figure}
\includegraphics[scale=0.75]{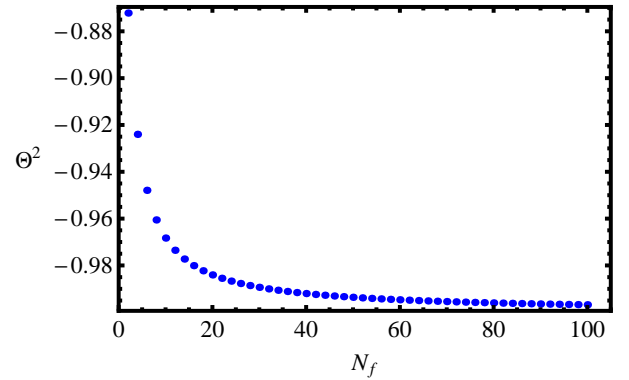}
\caption{\label{fig:theta2} 
First subleading irrelevant critical exponent $\Theta^{2}$ for $\Nf=2,4,6,\dots,100$ in $d=3$.}
\end{figure}

Let us now turn to the leading universal critical exponents
$\Theta^{1,2}$. The convergence of the polynomial expansion of the potential
is demonstrated in Tabs. \ref{tab1} \ref{tab2}, where the leading critical exponents
for $\Nf=2$ and $\Nf=12$ are listed for increasing truncation order. In each
case, the leading exponents converge to a stable value. We observe a
more rapid convergence for larger values of $\Nf$. The somewhat slower
convergence for more scalar dominated models is familiar from pure O($N$)
models. The leading two critical exponents are plotted as a function of  $\Nf$
in Figs.~\ref{fig:theta1} and \ref{fig:theta2}. Table \ref{tab3} lists the
leading exponents for increasing number of $\Nf$, illustrating the approach to
the analytical large-$\Nf$ results. 

\begin{table*}[h!t]\center
\begin{tabular}{p{50pt}||p{50pt}|p{50pt}|p{50pt}|p{50pt}|p{50pt}|p{50pt}p{0pt}}
\centering $2n$ & \centering $\Theta^{1}$ & \centering $\Theta^{2}$ & \centering $\Theta^{3}$ & \centering $\Theta^{4}$ & \centering $\Theta^{5}$ & \centering $\Theta^{6}$ & \\ \hline\hline
\centering $4$ & \centering $0.9928$ & \centering $-0.8687$ & \centering $-1.5743$ & \centering -- & \centering -- & \centering -- & \\
\centering $6$ & \centering $0.9766$ & \centering $-0.8743$ & \centering $-1.0624$ & \centering $-5.4313$ & \centering -- & \centering -- & \\
\centering $8$ & \centering $0.9831$ & \centering $-0.8721$ & \centering $-1.0790$ & \centering $-3.5622$ & \centering $-10.5959$ & \centering -- & \\
\centering $10$ & \centering $0.9821$ & \centering $-0.8720$ & \centering $-1.0999$ & \centering $-3.4194$ & \centering $-6.8111$ & \centering $-17.4807$ & \\
\centering $12$ & \centering $0.9819$ & \centering $-0.8723$ & \centering $-1.0897$ & \centering $-3.5628$ & \centering $-5.7949$ & \centering $-11.2156$ & \\
\centering $14$ & \centering $0.9821$ & \centering $-0.8722$ & \centering $-1.0911$ & \centering $-3.5104$ & \centering $-6.0610$ & \centering $-8.6408$ & \\
\centering $16$ & \centering $0.9820$ & \centering $-0.8722$ & \centering $-1.0920$ & \centering $-3.5062$ & \centering $-6.1190$ & \centering $-8.3598$ & \\
\centering $18$ & \centering $0.9820$ & \centering $-0.8722$ & \centering $-1.0914$ & \centering $-3.5202$ & \centering $-5.9849$ & \centering $-8.9972$ & \\
\centering $20$ & \centering $0.9821$ & \centering $-0.8722$ & \centering $-1.0915$ & \centering $-3.5132$ & \centering $-6.0516$ & \centering $-8.5869$ & \\
\centering $22$ & \centering $0.9821$ & \centering $-0.8722$ & \centering $-1.0916$ & \centering $-3.5135$ & \centering $-6.0514$ & \centering $-8.5820$ & \\
\end{tabular}
\caption{Non-Gau\ss ian critical exponents in $d=3$ for increasing polynomial
  truncations for $\Nf=2$. {The results for the critical exponent $\Theta^1$ agree within
the error bars with the result from Monte-Carlo (MC) simulations~\cite{Karkkainen:1993ef},
$1/\Theta^1_{\rm MC}=\nu_{\rm MC}\approx 1.00(4)$.}
}
\label{tab1}
\end{table*}

\begin{table*}[h!t]\center
\begin{tabular}{p{50pt}||p{50pt}|p{50pt}|p{50pt}|p{50pt}|p{50pt}|p{50pt}p{0pt}}
\centering $2n$ & \centering $\Theta^{1}$ & \centering $\Theta^{2}$ & \centering $\Theta^{3}$ & \centering $\Theta^{4}$ & \centering $\Theta^{5}$ & \centering $\Theta^{6}$ & \\ \hline\hline
\centering $4$ & \centering $0.9898$ & \centering $-0.9735$ & \centering $-1.0701$ & \centering -- & \centering -- & \centering -- & \\
\centering $6$ & \centering $0.9903$ & \centering $-0.9735$ & \centering $-1.0489$ & \centering $-3.2583$ & \centering -- & \centering -- & \\
\centering $8$ & \centering $0.9903$ & \centering $-0.9735$ & \centering $-1.0507$ & \centering $-3.1714$ & \centering $-5.5889$ & \centering -- & \\
\centering $10$ & \centering $0.9903$ & \centering $-0.9735$ & \centering $-1.0505$ & \centering $-3.1821$ & \centering $-5.3368$ & \centering $-8.1011$ &
\end{tabular}
\caption{Non-Gau\ss ian critical exponents in $d=3$ for increasing polynomial
  truncations for $\Nf=12$.}
\label{tab2}
\end{table*}

\begin{table*}[h!t]\center
\begin{tabular}{p{50pt}||p{50pt}|p{50pt}|p{50pt}|p{50pt}|p{50pt}|p{50pt}p{0pt}}
\centering $\Nf$ & \centering $\Theta^{1}$ & \centering $\Theta^{2}$ & \centering $\Theta^{3}$ & \centering $\Theta^{4}$ & \centering $\Theta^{5}$ & \centering $\Theta^{6}$ & \\ \hline\hline

\centering $2$ & \centering $0.9821$ & \centering $-0.8722$ & \centering $-1.0916$ & \centering $-3.5135$ & \centering $-6.0514$ & \centering $-8.5820$ & \\
\centering $4$ & \centering $0.9775$ & \centering $-0.9240$ & \centering $-1.1010$ & \centering $-3.3910$ & \centering $-5.7739$ & \centering $-8.2429$ & \\
\centering $12$ & \centering $0.9903$ & \centering $-0.9735$ & \centering $-1.0506$ & \centering $-3.1810$ & \centering $-5.3665$ & \centering $-7.6004$ & \\
\centering $50$ & \centering $0.9975$ & \centering $-0.9936$ & \centering $-1.0143$ & \centering $-3.0510$ & \centering $-5.1062$ & \centering $-7.1789$ & \\
\centering $100$ & \centering $0.9987$ & \centering $-0.9968$ & \centering $-1.0073$ & \centering $-3.0263$ & \centering $-5.0550$ & \centering $-7.0934$ & \\ \hline
\centering $\infty$ & \centering $1$ & \centering $-1$ & \centering $-1$ & \centering $-3$ & \centering $-5$ & \centering $-7$ & 
\end{tabular}
\caption{Non-Gau\ss ian critical exponents for various flavor numbers $\Nf$
  in $d=3$ in the $\rho^{11}$ truncation.}
\label{tab3}
\end{table*}

This approach is also visible in
the anomalous dimension and the fixed point values for the coupling, see
Tab.~\ref{tab4}. Whereas the fixed-point couplings are non-universal (holding
for the linear regulator in this case), the anomalous dimensions are universal
and illustrate the inequality \eqref{eq:ineq}.

Quantitatively, our results for the leading critical exponent can be
  compared to those studies of the quantum critical phase transition aiming at
  the long-range physics, as $\Theta^1$ is related to the correlation length
  exponent $\nu$ by $\nu=1/\Theta^1$. Together with the scalar anomalous
  dimension and corresponding scaling and hyperscaling relation, all
  thermodynamic exponents of the phase transition are determined.  Wherever
  comparable, our results agree quantitatively with the functional RG study of
  \cite{Hofling:2002hj} where both a polynomial expansion as well as a grid
  solution of the potential was used (note that our $\Nf$ counts
  four-component fermions, whereas \cite{Hofling:2002hj} uses two-component
  fermions). Also the agreement with results from other methods such as
  $1/\Nf$ expansions \cite{Gracey:1993kc} and Monte Carlo
  simulations \cite{Hands:1992be,Karkkainen:1993ef} is
  satisfactory. Discrepancies are mainly visible only in the anomalous
  dimensions for small $\Nf$, a feature familiar from scalar
  models. Also the polynomial expansion of the potential converges somewhat
  slower for the special case $\Nf=1$, where our results for the leading
  exponents are compatible with those of \cite{Hofling:2002hj}, but subleading
  exponents seem to require a precise solution of the potential flow.  

To summarize, the nonperturbative features of the Gross-Neveu model near
  criticality can well be described by the functional renormalization group,
  as the model interpolates between two well-accessible limits: the
  large-$\Nf$ limit for $\Nf\to\infty$ and the Wilson-Fisher fixed point in
  the Ising universality class for $\Nf\to 0$. 
{Our results suggest that the Gross-Neveu model is asymptotically safe
for all $\Nf>0$ and that the model depends on only one parameter, even
when we take into account corrections beyond the large-$\Nf$ limit. This
might {provide} helpful information{, e.g.,} for a systematic study of the finite-temperature 
phase diagram of the Gross-Neveu model beyond the large-$\Nf$ approximation.
}

\begin{table*}[h!t]\center
\begin{tabular}{p{50pt}||p{50pt}|p{50pt}|p{50pt}|p{50pt}p{0pt}}
\centering $\Nf$ & \centering $\eta_{\sigma,\ast}$ & \centering $\eta_{\psi,\ast}$ & \centering $h_{\ast}^{2}$ & \centering $\lambda_{2,\ast}$ & \\ \hline\hline
\centering $2$ & \centering $0.7596$ & \centering $0.0320$ & \centering $4.5565$ & \centering $0.3956$ & \\ 
\centering $4$ & \centering $0.8870$ & \centering $0.0138$ & \centering $5.2820$ & \centering $0.5846$ & \\ 
\centering $12$ & \centering $0.9644$ & \centering $0.0041$ & \centering $5.7206$ & \centering $0.7263$ & \\
\centering $50$ & \centering $0.9917$ & \centering $0.0009$ & \centering $5.8746$ & \centering $0.7822$ & \\
\centering $100$ & \centering $0.9958$ & \centering $0.0005$ & \centering $5.8983$ & \centering $0.7911$ & \\ \hline 
\centering $\infty$ & \centering 1 & \centering 0 & \centering $5.9218$ & \centering $0.8$  
\end{tabular}
\caption{Non-Gau\ss ian fixed-point values of the universal anomalous dimensions
  for various flavor numbers $\Nf$ in $d=3$. The (non-universal) fixed-point couplings hold for the linear
  regulator. {In Monte-Carlo (MC) simulations~\cite{Karkkainen:1993ef} $\eta_{\sigma,\ast}^{\rm MC}=0.754(8)$
has been found for $\Nf=4$ two-component fermions (corresponding to $\Nf=2$ in our language).}
} 
\label{tab4}
\end{table*}

\section{Conclusions and Outlook}

We have used the functional RG to describe the UV behavior of the Gross-Neveu
model in $2<d<4$ dimensions. In agreement with many earlier results in the
literature, the model is nonperturbatively renormalizable by means of a
non-Gau\ss ian fixed point, providing a simple example of asymptotic safety in
Weinberg's sense. The perturbative conclusion about
nonrenormalizability is a mere artifact of naive power-counting which is only
justified near the Gau\ss ian fixed point.

In this work, we have summarized these conclusions in the functional RG
language as it is also extensively used recently to explore the possibility of
quantizing gravity within pure quantum field theory. This pedagogic character
of our work is amended by new results in the large-$\Nf$ limit, where
the fixed-point potential as well as all critical exponents can be computed
analytically. Moreover, we have also provided finite $\Nf$ results,
demonstrating that the model interpolates between the large $\Nf$ limit and
a purely bosonic model in the Ising universality class. 

{
Some final comments are in order: the simple fermionic Gross-Neveu action is
minimalistic in the sense that it suffices to put the system into the right
``Gross-Neveu universality class''. Comparing the fermionic action with the
fixed-point action in the partially bosonized version, we have to conclude
that the fixed-point action in the fermionic language is far more complicated
{than the simple ansatz~\eqref{eq:eff_action_fermion}}. 
In general, it will contain higher-order
as well as non-local interaction terms. The deviations from the simple
Gross-Neveu structure, however, are irrelevant operators which do not modify
the predictive power of the Gross-Neveu model.} 

{
The simple Gross-Neveu action is also incomplete in the sense that it does not
exhibit all possible four-fermi terms compatible with the defining symmetries
of the model. For instance, a Thirring-like interaction $~(\bar\psi \gamma_\mu
\psi)^2$ is also invariant under the symmetries of the Gross-Neveu model and
will thus generically be generated by the RG flow. A more complete RG analysis
thus has to include these terms facilitating the appearance of further
fixed-points, as is known for the Thirring model \cite{Gies:2010st}. If so,
this implies the possible existence of further UV completions of fermionic
models with the same symmetries as the Gross-Neveu model. }

{Finally, let as mention once more that the property of asymptotic safety
  in the Gross-Neveu model is tightly related to the occurrence of a quantum
  phase transition of 2nd order separating a disordered phase from a phase
  with broken (discrete) chiral symmetry. More generally, models with such
  2nd-order quantum phase transitions are guaranteed to be asymptotically
  safe. Whether the converse is true, i.e., whether asymptotically safe models
  always exhibit a physically relevant order-disorder quantum phase
  transition, is an interesting question for future studies.}

\acknowledgments The authors thank L. Janssen and M. M. Scherer for useful discussions and
acknowledge support by the DFG under grants Gi~328/1-4, Gi~328/5-1 (Heisenberg
program), GRK1523 and FOR 723.

\appendix 

\section{Threshold functions}\label{App:tf}
The regulator dependence of the flow equations is encoded in dimensionless threshold
functions which arise from 1PI diagrams incorporating bosonic and/or fermionic fields.
In this work we have employed a linear regulator~\cite{Litim:2000ci,Litim:2001up,Litim:2001fd},
see Eqs.~\eqref{eq:fermReg} and~\eqref{eq:bosReg}.
For the threshold functions appearing in the flow equations 
in Sect.~\ref{sec:bos}, we then find
\be
&&l_0^d(\omega;\eta_{\sigma})= \frac{2}{d}\left(1-\frac{\eta_{\sigma}}{d+2}\right)\frac{1}{1+\omega}\,,\nn
\ee
\be
&&l_0^{(F)d}(\omega;\eta_{\psi})=\frac{2}{d}\left(1-\frac{\eta_{\psi}}{d+1}\right)\frac{1}{1+\omega}\,,\nn
\ee
\be
&&l_{1,1}^{(FB)d}(\omega_1,\omega_2;\eta_{\psi},\eta_{\sigma})=\frac{2}{d}\frac{1}{(1+\omega_1)(1+\omega_2)}\times\nn\\
&&\quad \times\left\{\left(1-\frac{\eta_{\psi}}{d+1} \right)\frac{1}{1+\omega_1} + \left(1-\frac{\eta_{\sigma}}{d+2} \right)\frac{1}{1+\omega_2}
\right\}\,,\nn
\ee
\be
m_4^{(F)d}(\omega;\eta_{\psi})&=&\frac{1}{(1+\omega)^4}+\frac{1-\eta_{\psi}}{d-2}\frac{1}{(1+\omega)^3}\nn\\
&&\quad-\left(\frac{1-\eta_{\psi}}{2d-4} +\frac{1}{4}\right)\frac{1}{(1+\omega)^2}\,,\nn
\ee
\be
&& m_{1,2}^{(FB)d}(\omega_1,\omega_2;\eta_{\psi},\eta_{\sigma})\nn\\
&&\qquad\qquad =\left(1-\frac{\eta_{\sigma}}{d+1}\right)\frac{1}{(1+\omega_1)(1+\omega_2)^2}\,.\nn
\ee
Note that the linear regulators~\eqref{eq:fermReg} and~\eqref{eq:bosReg} render $m_{1,2}^{(FB)d}$ 
independent of $\eta_{\psi}$.

\end{document}